\documentclass{aa}
\usepackage{graphicx}
\usepackage{txfonts}
\usepackage{natbib}
\newcommand{\beq}   {\begin{equation}}
\newcommand{\eeq}   {\end{equation}}
\newcommand{\kms}   {km~s$^{-1}$}
\newcommand{\water}   {H$_2$O~}
\def\gs{\mathrel{\raise0.35ex\hbox{$\scriptstyle >$}\kern-0.6em
\lower0.40ex\hbox{{$\scriptstyle \sim$}}}}
\def\ls{\mathrel{\raise0.35ex\hbox{$\scriptstyle <$}\kern-0.6em
\lower0.40ex\hbox{{$\scriptstyle \sim$}}}}

\begin{document}
   \title{The magnetic field of the proto-planetary nebula candidate IRAS 19296+2227}

   \titlerunning{The magnetic field of IRAS 19296+2227}


   \author{W.H.T. Vlemmings\inst{1}\and
           H.J. van Langevelde\inst{2,3}
          }

   \offprints{WV (wouter@astro.uni-bonn.de)}

    \institute{Argelander Institute for Astronomy, University of Bonn,
     Auf dem H{\"u}gel 71, 53121 Bonn, Germany
         \and
         Joint Institute for VLBI in Europe, Postbus 2, 
                7990~AA Dwingeloo, The Netherlands
         \and
         Sterrewacht Leiden, Leiden University, Niels Bohrweg 2, 
                2333 CA Leiden, The Netherlands    
                          }

   \date{Received ; accepted }

   \abstract{Magnetic fields are thought to be one of the possible mechanisms responsible for shaping the generally spherical outflow of evolved stars into often aspherical planetary nebulae. However, direct measurements of magnetic fields during the transition to the planetary nebula phase are rare.}{The aim of this project is to expand the number of magnetic field measurements of stars in the (proto-)planetary nebula phase and find if the magnetic field strength is sufficient to affect the stellar outflow.}{We used Very Long Baseline Array observations to measure the circular polarization due to the Zeeman splitting of 22~GHz \water masers in the envelope of the proto-planetary nebula candidate star IRAS~19296+2227 and the planetary nebula K3-35.}{A strong magnetic field of $B_{\rm ||}=-135\pm28$ is detected in the \water maser region of the proto-planetary nebula candidate IRAS 19296+2227. The \water masers of K3-35 are too weak to detect circular polarization although we do present the measurements of weak linear polarization in those masers.}{The field measured in the masers of IRAS 19296+2227 is dynamically important and, if it is representative of the large scale field, is an important factor in driving the stellar mass loss and shaping the stellar outflow. 
 }

   \maketitle

\section{Introduction}

At the end of their evolution, a majority of stars go through a period
of high mass loss that is an important source for replenishing
interstellar space with processed materials. During the asymptotic
giant branch (AGB) phase, this mass loss produces circumstellar
envelopes which are found to undergo a major modification during the
rapid transition from AGB star to Planetary Nebula (PN). The standard
assumption is that the initial slow AGB mass loss in a short time
changes into a fast superwind generating shocks and accelerating the
surrounding envelope \citep{Kwok78}. However, a large fraction of PNe
have asymmetric shapes, with the majority of the young PNe being
bipolar. Thus, at some point during the evolution to a PNe the AGB
stars must undergo a process in which the spherically symmetric
outflow is altered to produce a-spherical PNe morphologies. It has
been shown that the energy contained in the outflows of young bipolar
PNe is often orders of magnitude larger than can be provided by
radiation pressure \citep{Bujarrabal01}. The source of this energy has
been argued to be magnetic fields, binary or disk interaction or a
combination of these \citep[see e.g.][and references
therein]{Balick02, Frank07}.

Maser polarization observations are the predominant source of
information about the role of magnetic fields during the late stages
of stellar evolution. Most observations have focused on the masers in
the envelopes of AGB stars, as OH, H$_2$O and SiO masers are fairly
common in these sources, and have revealed strong magnetic fields
throughout the entire envelope \citep[e.g.][]{Etoka04, Vlemmings05,
  Herpin06}. Because they are a short-lived ($\approx1000$~yr)
transition phase between AGB star and PNe, proto-planetary nebulae
(PPNe) are fairly rare. However, OH maser observations reveal similar
strength magnetic fields as in their progenitor AGB stars
\citep[e.g][]{Bains03, Bains04}. Additionally, a very small fraction
of the PPNe maser stars show highly collimated H$_2$O maser jets. 
  Currently 11 of these so-called 'water-fountain' sources have been
  identified, and they are likely the progenitor of bipolar PNe
  \citep[][and references therein]{Imai07}. Polarization observations
of the \water masers of W43A have recently revealed that the maser jet
is magnetically collimated \citep{Vlemmings06c}.

Here we present the detection of a strong magnetic field in the \water
maser region of the circumstellar envelope of the PPNe candidate
IRAS~19296+2227, only the second such detection in an object beyond
the AGB phase. Additionally, we shortly describe the measurement of
linear polarization of the \water masers of the PNe K3-35
\citep{Miranda01}, which was observed together with
IRAS~19296+2227. The PNe K3-35 is only one of 4 PNe towards which
  \water masers have been detected \citep{Suarez07}. IRAS~19296+2227
is a source with both \water and OH maser emission and IRAS colors
indicative of an AGB star with strong mass loss. Its near kinematic
distance is estimated by \citet{Engels96} to be $3.2\pm1.3$~kpc, which
implies an infrared luminosity of $5000~L_\cdot$. Although this
luminosity is consistent with that of an AGB star, its maser
properties are atypical since both the OH and \water spectra are
narrow compared with the maser spectra of OH/IR stars. However, its
maser properties do resemble those of a number of PPNe
candidates. Therefore, \citet{Engels96} concluded that IRAS~19296+2227
should also be considered a PPNe candidate.

\begin{figure*}[htf]
   \resizebox{\hsize}{!}{\includegraphics{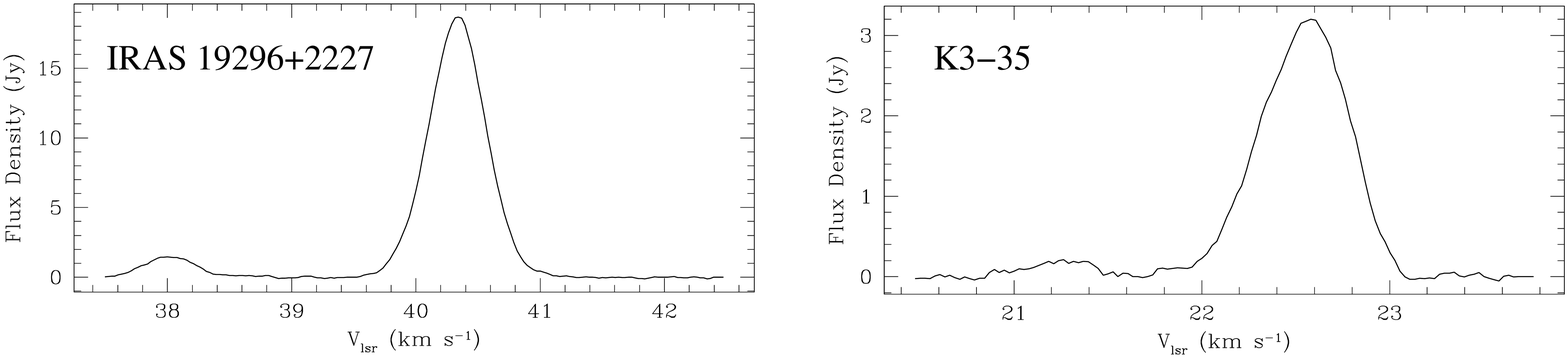}} \hfill
\caption[specs]{Observed VLBA spectra as measured in the image plane of the \water masers of IRAS~19296+2227 (left) and K3-35 (right).}
\label{Fig:specs}
\end{figure*}

\section{Observations}
\label{obs}

The NRAO\footnote{The National Radio Astronomy Observatory (NRAO) is a
  facility of the National Science Foundation operated under
  cooperative agreement by Associated Universities, Inc.} Very Long
Baseline Array (VLBA) was used on June 20 2003 to observe the \water
masers of the PNe K3-35 and PPNe candidate IRAS 19296+2227. The
average beam width is $\approx 1.0 \times 0.3$~mas and $1.0 \times
0.7$~mas at the frequency of the $6_{16} - 5_{23}$ rotational
transition of H$_2$O, 22.235 GHz for IRAS 19296+2227 and K3-35
respectively. We used 4 baseband filters of 1 MHz width, which were
overlapped to get a velocity coverage of $\approx 44$~km/s. However,
for both sources maser emission was only detected in one of the
observing bands. The data were correlated multiple times. The initial
correlation was performed with modest ($7.8$~kHz$ = 0.1$~\kms)
spectral resolution, which enabled us to generate all 4 polarization
combinations (RR, LL, RL and LR). Two additional correlator runs were
performed with high spectral resolution ($1.95$~kHz$ = 0.027$~\kms)
which therefore only contained the two polarization combinations RR
and LL. Such high resolution is needed to detect the circular
polarization pattern of the \water masers at sufficient accuracy. Each
source-calibrator pair was observed for 6 hours. The calibrator was
observed for $1.5$ hours in a number of scans equally distributed over
the 6 hours.

\subsection{Calibration}

The data analysis path is described in detail in
\citet{Vlemmings02}. It follows the method of \citet{KDC95} and was
performed in the Astronomical Image Processing Software package
(AIPS). The calibration steps were performed on the data-set with
modest spectral resolution. Delay and bandpass calibration were done
using the calibrator sources J1850+2825 and 3C84, while fringe fitting and
self-calibration were performed on a strong maser feature. The
calibration solutions were then copied and applied to the high
spectral resolution data-set.  Finally, corrections were made for
instrumental feed polarization using the unpolarized calibrator
3C84. Unfortunately, no good calibrator data was available for the
polarization angle calibration, making it impossible to properly
determine the direction of the linear polarization vectors.

The resulting noise in emission free channels of the modest spectral
resolution I,Q, and U data cubes was $\sim4.0$~mJy~Beam$^{-1}$ while
that in the high spectral resolution I and V cubes was
$\sim8.2$~mJy~Beam$^{-1}$. In the channels with strong maser emission,
the noise increased by $\sim10\%$.

Fig.~\ref{Fig:specs} shows the \water maser spectra for
IRAS~19296+2227 and K3-35 as measured in the image plane. The
high-spectral resolution stokes I image cubes were analyzed using the
{\it AIPS} task SAD, which was used to fit all maser features with
peak flux densities higher than 8 times the rms in the channel map
with two-dimensional Gaussian components. Subsequently, we only
retained components that were detected in at least 10 consecutive
channels, corresponding in total to $\sim0.27$~\kms, as narrower features are
unlikely to be real. To enable the most accurate positional comparison
with the observations in \citet{Marvel99} we then also determined the
mean right ascension and declination off-sets with respect to the
reference feature using a flux density-squared weighting scheme. The
peak flux, full-width half-maximum ($\Delta v_L$) and velocity
($V_{\rm LSR}$) were determined by fitting a Gaussian to the feature
spectra.

\section{Results}
\label{results}

\begin{table*}
\caption{Results}
\begin{tabular}{|l|c|c|c|c|c|c|c|c|}
\hline
Name & Feature  & Flux (I) & $V_{\rm rad}$ & $\Delta
v_{\rm L}$ & $\Delta\alpha^{a}$ & $\Delta\delta^{a}$ & $B_{\rm ||}$ & $m_l$ \\
 & & (Jy beam$^{-1}$) & (\kms) & (\kms) & (mas) & (mas) & (mG) & $(\%)$\\
\hline
\hline
IRAS 19296+2227 & a & 0.54 & 38.04 & 0.41 & 6.18 & -11.07 & & \\  
                & b & 0.11 & 38.64 & 0.38 & 3.49 & -6.37 & & \\
                & c & 0.08 & 40.07 & 0.51 & -16.73 & 3.86 & & \\
                & d & 1.70 & 40.28 & 0.59 & 0.88 & -0.30 & & \\
                & e & 0.24 & 40.30 & 0.46 & 0.46 & -1.12 & & \\
                & f & 8.13 & 40.33 & 0.48 & 0.00 & 0.00 & -135$\pm$28 & \\  
                & g & 0.07 & 40.34 & 0.33 & -1.55 & -2.23 & & \\
                & h & 0.18 & 40.75 & 0.64 & -16.04 & 4.99 & & \\
\hline
K3-35 & a & 0.22 & 21.26 & 0.40 & 14.00 & 8.70 & & \\
      & b & 0.23 & 22.16 & 0.33 & 5.54 & 3.55 & & $1.6\pm0.2$ \\
      & c & 3.47 & 22.58 & 0.50 & 0.00 & 0.00 & & \\
\hline
\multicolumn{9}{l}{$^{a}$ Positional offsets measured from the maser reference features IRAS~19296+2227$f$ and K3-35$c$.}
\end{tabular}
\label{Table:results}
\end{table*}

\begin{figure}
   \resizebox{\hsize}{!}{\includegraphics{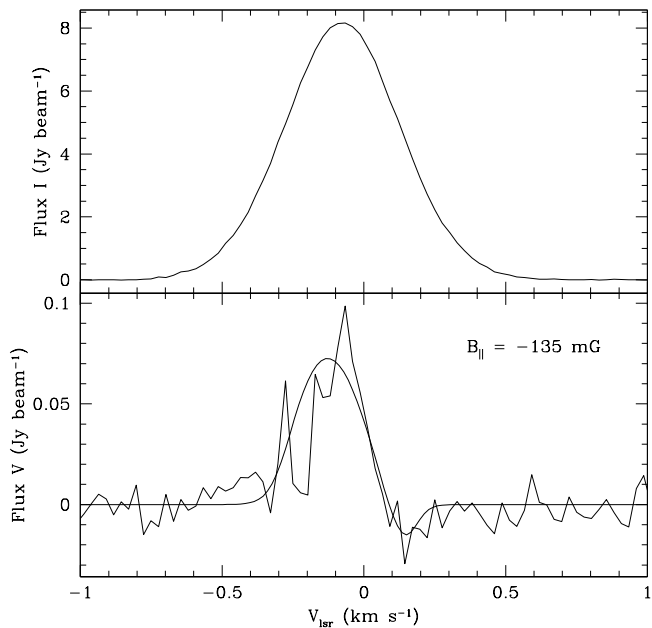}}
   \hfill
   \caption[magfield]{Total power ($I$) and circular polarization
     ($V$)-spectrum for the strongest maser feature ($f$) of
     IRAS~19296+2227. The thick solid line in the bottom panel shows
     the best model fit to $V$. This fit corresponds to a magnetic
     field strength of $-135$~mG along the maser line-of-sight.}
\label{Fig:magfield}
\end{figure}

\begin{figure*}[htf]
   \resizebox{\hsize}{!}{\includegraphics{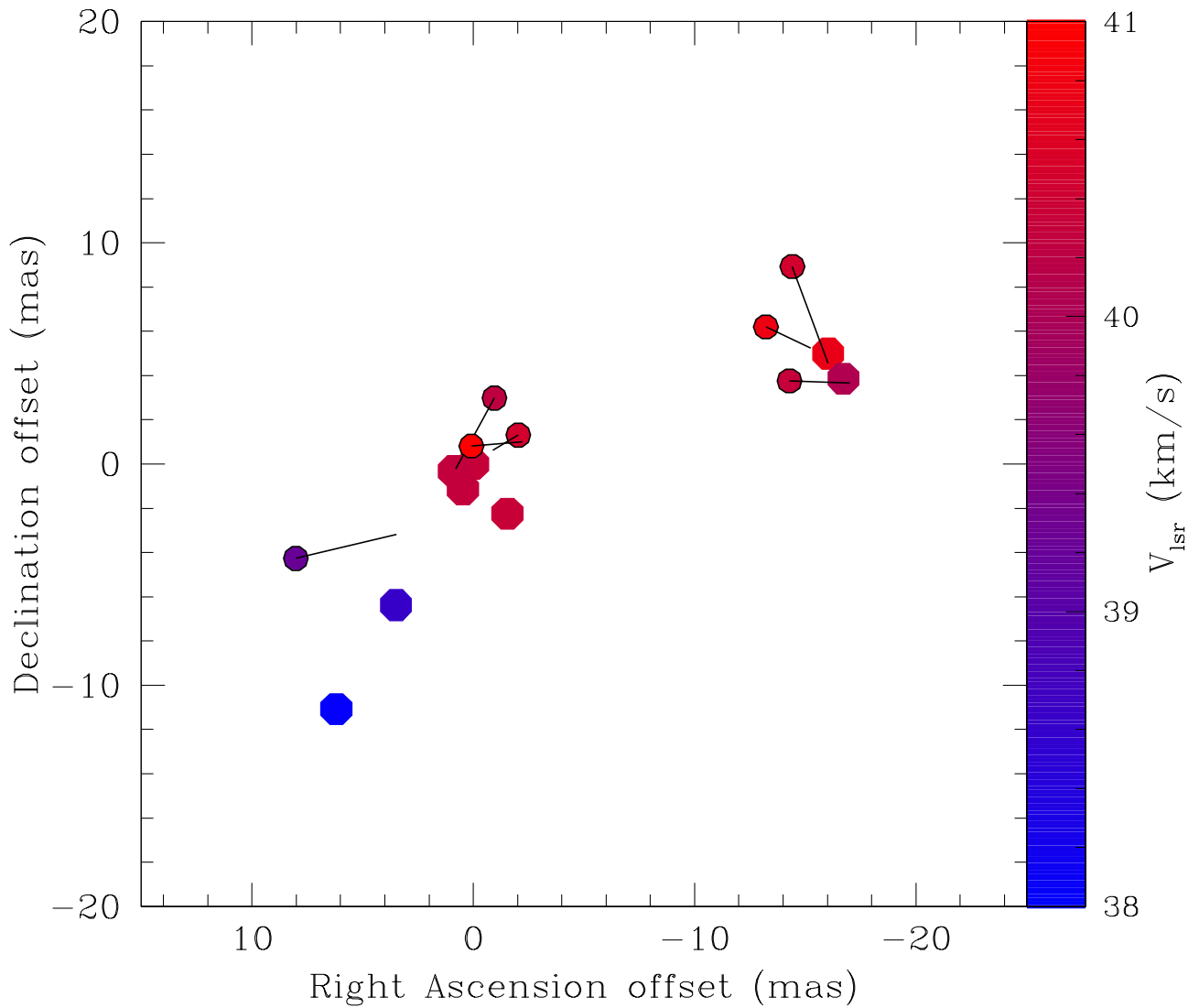}} \hfill
   \caption[map]{The \water maser region of IRAS~19296+2227. The large
     filled octagons are the maser features detected in our
     observations and are color-coded according to LSR velocity. The
     smaller circles with black edges and vectors denote the predicted
     positions when transposing the maser features with proper motion
     determinations from \citet{Marvel99} between May 31 1997 and Jun
     20 2003. The vectors indicate the maser feature trajectory,
     ending at the 1997 positions. This illustrates that the large scale
     maser proper motions are significantly less than determined in
     \citet{Marvel99}.}
\label{Fig:map}
\end{figure*}

The result of our analysis of the \water masers features around
IRAS~19296+2227 and K3-35 is presented in
Table.~\ref{Table:results}. Here we list the feature LSR velocity,
full-width half-maximum ($\Delta v_L$) and the right ascension and
declination positional offsets from the maser reference features
($\Delta\alpha$ and $\Delta\delta$). A map of the maser distribution
of IRAS~19296+2227 is shown in Fig.~\ref{Fig:map}. Circular
polarization of $1.1\pm0.2$\% was detected for the strongest maser ($f$)
of IRAS~19296+2227 and its spectrum is shown in
Fig.~\ref{Fig:magfield}. No circular polarization was detected for the
weaker features of K3-35. On the other hand, linear polarization at a
level of $1.6\pm0.2$\% was only detected on feature $b$ of K3-35 while
no significant linear polarization was detected for IRAS~19296+2227.

The relation between circular polarization fraction $P_V$,
  magnetic field strength along the line of sight $B_{\rm ||}$ and
  maser line width $\Delta v_{\rm L}$ is given by $P_V=2 A_{\rm FF'}
  B_{\rm ||} / \Delta v_{\rm L}$. However, the coefficient $A_{\rm
    FF'}$ is a function of maser saturation level and for our analysis
  of the circular polarization of IRAS~19296+2227 $f$ we used the full
  radiative transfer non-LTE interpretation described in
  \citet{Vlemmings02} to take this into account. In this
interpretation, the coupled equations of state of the three dominant
hyperfine components of the 22~GHz \water maser transition was solved,
following \citet{NW92}, for a linear maser in the presence of a
magnetic field. It is then possible to fit the total intensity and
circular polarization spectra directly to model spectra for different
intrinsic maser thermal line widths ($\Delta v_{\rm th}$). The
emerging maser flux densities of the model spectra are expressed in
$T_b\Delta\Omega$, where $T_b$ is the brightness temperature (in $K$)
and $\Delta\Omega$ the beaming solid angle (in ${\rm sr}$). The
uncertainties in these fits are further discussed in
\citet{Vlemmings06b}. The best fit model to the IRAS~19296+2227
spectrum gives $B_{\rm ||}=-135\pm28$~mG, $\Delta v_{\rm
  th}=1.0\pm0.3$~\kms and ${\rm log}(T_b\Delta\Omega)=9.4\pm0.4$.

\section{Discussion}
\label{disc}

\subsection{The \water maser shell of IRAS 19296+2227}

The distribution of the \water maser features around IRAS~19296+2227
is shown in Fig.~\ref{Fig:map}. The masers lie in an elongated
structure with a length of $\sim40$~mas. At a distance of 3.2~kpc,
this corresponds to a linear size of $\sim130$~AU. As no larger
structure was found in VLA observations \citep{Marvel99}, this is
likely the full extent of the \water maser shell.

\citet{Marvel99} published the first VLBA images made at 3 epochs of
observations in 1997 (May 31, June 17 and July 10). Comparing our data
with the earlier observations, we find that, while the flux has
decreased by more than a factor of two, the maser distribution is
extremely similar. Our observations fail to detect several of the weak
features at $V_{\rm LSR}>41$~\kms but the two most prominent maser
groups identified in \citet{Marvel99} have persisted for over 6
years. Still, the center velocity of some of the maser features, such
as the strongest reference feature, has drifted by $\approx0.2$~\kms,
making it impossible to match up individual maser features. As a
result, direct proper motion measurements cannot be made. If the maser
motions detected by \citet{Marvel99} are the genuine large scale
motions, one expects, after 6 years of motion, a significantly
different maser distribution.  Fig.~\ref{Fig:map} indicates the
positions of the maser features for which proper motions were
determined in \citet{Marvel99}, transposed over 6 years.  Assuming the
maser reference feature was the same in all observational epochs, the
predicted maser distribution fails to match our observed distribution,
implying that the proper motions determined in \citet{Marvel99} are
likely dominated by random motions of the maser features. Assuming the
maser features $c$ and $h$ in our observations originate from the
maser clump located at $-16.1$ and $4.5$~mas Right Ascension and
Declination off-set respectively in epoch 1 of \citet{Marvel99}, the
maximum proper motion of these features is $\sim0.15$~mas~yr$^{-1}$
which corresponds to $\sim2.3$~\kms.


As described above, the combined analysis of the total intensity and
circular polarization spectra provides additional information on the
intrinsic conditions in the maser region. The best fit intrinsic
thermal line width of the material that gives rise to the
IRAS~19296+2227 \water masers was found to be $\Delta v_{\rm
  th}=1.0$~\kms.  Since $\Delta v_{\rm th}\approx0.5(T/100)^{1/2}$,
with $T$ the temperature in the maser region, the \water masers in the
envelope of IRAS~19296+2227 exist at $\sim400$~K. Additionally, we
find that the emerging brightness temperature of the strongest maser
feature is $T_b\Delta\Omega\approx2.5\times10^{9}$~K~sr. Because the
maser is marginally resolved we take the size of the maser to be
$0.3$~mas. The measured brightness temperature of IRAS~19296+2227$f$
is thus $T_b\approx2\times10^{11}$~K. This implies a beaming solid
angle $\Delta\Omega\approx1.2\times10^{-2}$~sr, which is typical for the
beaming angle of the \water masers around evolved stars
\citep{Vlemmings05b}. Finally, as detailed in \citet{Vlemmings06b},
the emerging brightness temperature implies that the masers are
unsaturated.

\subsection{Linear polarization}

No linear polarization was detected from the IRAS 19296+2227 \water
masers down to $\sim0.4\%$. This is a similar limit as found for the
Mira and supergiant stars in our previous sample \citep{Vlemmings02,
  Vlemmings05}. However, the strongest \water maser of the PNe K3-35
was linearly polarized at a level of $\sim1.6\%$. Unfortunately, the
lack of good polarization angle calibration made it impossible to
determine the polarization vector orientation. The fractional
polarization of the K3-35 maser is similar to that found for
\water masers in star forming regions and those in the
'water-fountain' PPNe source W43A
\citep{Vlemmings06b,Vlemmings06d}. Considering that the maser linear
polarization fraction increases with maser saturation
\citep[e.g][]{Deguchi90}, this seems to indicate that the saturation
level of the \water masers in the evolved stellar envelopes is on
average less than that of the maser in star forming regions, PNe and
related 'water-fountain' sources.

\subsection{The magnetic field of IRAS 19296+2227}

The circular polarization of the strongest \water maser feature ($f$)
of IRAS 19296+2227 implies a magnetic field of $B_{\rm
  ||}=-135\pm28$~mG in the \water maser region. The field strength is
similar to the fields previously detected using \water maser
polarization observations in the envelopes of evolved stars
\citep{Vlemmings02, Vlemmings05, Vlemmings06c}. 

The origin of the magnetic field is unclear. As only one maser feature
was strong enough to detect a magnetic field, a field local to the
masers cannot be ruled out \citep[e.g][]{S02}. However,
\water, OH and SiO maser polarization observations of a number of
other evolved stellar sources reveal ordered large scale magnetic
fields \citep[e.g.][]{Bains03, Bains04, Cotton06, KD97, Vlemmings05,
  Vlemmings06c}. 

A field of $135$~mG in a region with $T\approx400$~K and a hydrogen
number density $n_{\rm H_2}$ of $\sim10^9$~cm$^{-3}$, typical for
\water masers, implies that the ratio between magnetic and thermal
pressure $B^2/(8\pi n_{\rm H_2} k T)\approx15$. Assuming the masing
material has a velocity of $\approx3$~\kms, the ratio between magnetic
and dynamic pressure $B^2/(8\pi \rho v^2)\approx5$, where $\rho$ and
$v$ are the density and velocity of the maser medium.  Thus, if the
measured field strength in the IRAS~19296+2227 \water maser region
reflects a larger scale field, the magnetic field is strong
enough to play an important role in the dynamics of the stellar
outflow and possibly also in driving the mass-loss through the effects
of Alfv{\'e}n waves near the stellar surface \citep[e.g.][]{Falceta02,
  Vidotto06}.

\section{Conclusions}
\label{concl}
We have measured a magnetic field of $B_{\rm ||}=-135\pm28$~mG in the
\water maser region of the PPNe candidate star IRAS 19296+2227. The
field strength is of the same order as the previously measured field
strengths in evolved stellar envelopes. This implies that there is no
significant evolution of the magnetic field in the transition from AGB
star to early PNe. The magnetic field, if representative of the
  large scale field, is strong enough to play an important role in the
  dynamics of the stellar outflow and possibly also in driving the
  mass-loss. However a direct interpolation of the measured field
  strength to the stellar atmosphere is impossible.

While no linearly polarized \water maser emission is detected in the
envelope of IRAS~19296+2227, $\sim 1.6$\% fractional linear
polarization is found for the strongest maser feature of the PNe
K3-35, implying that the masers aree at least partially
saturated. With sufficiently sensitive observations, the measurement
of linear polarization of the \water masers of PNe will enable us to
probe the magnetic field morphology early in the PNe phase. The
circular polarization measurement and lack of linear polarization
suggests the \water masers of IRAS~19296+2227 are unsaturated.


\end{document}